\documentstyle[12pt]{article}

\newcommand{\be}{\begin{equation}}
\newcommand{\ee}{\end{equation}}
\newcommand{\beq}{\begin{eqnarray}}
\newcommand{\eeq}{\end{eqnarray}}

\newcommand{\tD}{{\tilde D}}
\newcommand{\tX}{{\tilde X}}
\newcommand{\tmu}{{\mu'}}
\newcommand{\tl}{\lambda'}
\newcommand{\tln}{\lambda_n'}
\newcommand{\w}{{\omega}}
\newcommand{\K}{{\cal K}}
\newcommand{\CD}{{\cal D}}
\newcommand{\CZ}{{\cal Z}}
\newcommand{\CR}{{\cal R}}
\newcommand{\Z}{{\bf Z}}

\newcommand{\tr}{{\rm tr}}
\newcommand{\p}{{\partial}}
\newcommand{\hf}{\frac12}
\newcommand{\np}{{Nucl. Phys. }}
\newcommand{\pl}{{Phys. Lett. }}

\begin{document}

\title{ \Large \bf
Correlators in 2D string theory with vortex condensation}
\author{\normalsize 
S.~Yu.~Alexandrov$^{1,2}$\protect\thanks{
e.mail: alexand@spht.saclay.cea.fr.
 Also at V.A.~Fock Department of Theoretical Physics, St.~Petersburg
University, Russia} \
and  V.~A.~Kazakov$^2$\protect\thanks{
e.mail: kazakov@physique.ens.fr}}

\date{}
\maketitle

\vspace{-0.9cm}
\begin{center}
\it
$^1$Service de Physique Th\'eorique, C.E.A. - Saclay, 91191 Gif-sur-Yvette
CEDEX, France\\
\ \\
$^2$Laboratoire de Physique Th\'eorique de l'\'Ecole Normale
Sup\'erieure\footnote[0]{
Unit\'e de Recherche du
Centre National de la Recherche Scientifique et de  l'Ecole Normale
Sup\'erieure et \`a
l'Universit\'e de Paris-Sud.}\\
24 rue Lhomond, 75231 Paris Cedex 05, France
\end{center}

\hspace{10.1cm} LPTENS-01/18, SPhT-t01/036

\vspace{0.2cm}

\begin{abstract}
We calculate one- and two-point correlators of winding operators in 
the matrix model of 2D string theory compactified on a circle,
recently proposed for the description of string dynamics 
on the 2D black hole background. 
\end{abstract}

\section{ Introduction }

The matrix quantum mechanics (MQM) approach to the 2D string theory
has been proven very effective, in particular, in the case of the
compactification on a circle of a radius $R$ \cite{GRKL,BULKA}.
Recently, it was shown \cite{KKK} that this approach   is
capable to capture such important phenomena as the emergence of a
black hole background and condensation of winding modes (vortices on
the world sheet).

From the statistical-mechanical point of view such a model describes
a system of planar rotators ($XY$-model) on a two-dimensional
surface with  fluctuating metric.  In this sense it is a natural
generalization of its analogue on the flat 2D space studied long ago
by Berezinski, Kosterlitz and Thouless (BKT).
The role of the fluctuating surfaces is played by the planar graphs
of the MQM and the vortices appear due to the compactification of the
time direction.

As it is usual for the matrix model approach, it enables us to apply
the powerful framework of classical integrability to the computations
of physical quantities, such as partition functions of various genera
or correlators of physical operators.

The appropriate MQM model for the compactified 2D string is the
inverted matrix oscillator with twisted boundary conditions
\cite{BULKA}. The partition function of such a system as a function of
couplings of the potential for the twist angles appears to be a
$\tau$-function of the Toda lattice hierarchy \cite{HKK,KKK}. As such,
it satisfies an infinite set of partial difference-differential Hirota
equations. The couplings play the role of ``times'' of commuting flows
in the Toda hierarchy. 
As was shown in \cite{KKK} they are at the same time
the couplings of vortex operators of various vorticity charges. 
The tree approximation for the string theory
corresponds to the dispersionless limit of Toda equations.

In this paper we will use Hirota equations to calculate all one-point
and two-point correlators of vortex operators with arbitrary vorticity
charges for the compactified 2D string in the tree-like approximation
in the regime of the radius of compactification $\frac12
R_{KT}<R<R_{KT}$ ($R_{KT}$ is the BKT radius), when condensation of
vortices of the lowest charge $n=\pm 1$ takes place.  Unlike the
standard BKT model in flat 2D space, where the Debay screening in the
vortex plasma destroys the long range correlations, the BKT vortices
in the fluctuating 2D metric can be described as a conformal matter
with central charge $c=1$ coupled to the metric.

An important particular case of this model, for $R={3\over 4}R_{KT}$, 
claimed in \cite{KKK} to describe the
2D string in the dilatonic black hole background, due to the
conjecture of \cite{FZZ} based on the results of \cite{FAT,FB}.  This
conjecture establishes a kind of a weak-strong duality between two
CFT's: the dilatonic black hole and the so called Sine-Liouville
theory. We will try to compare our results with a few calculations of
correlators in the continuum CFT performed in \cite{FAT,FB,FZZ} and
discuss the difficulties of the direct comparison. 

In the next section we briefly describe the basic ideas of the MQM
approach to the 2D string theory with vortex excitations; the
Toda integrable structure of the model allows to formulate  
the  Hirota equations describing the dynamics of vortices. 
In section 3, using Hirota equations the two point correlators of
arbitrary vorticities are expressed through the one point
correlators.
In section 4 the one point correlators are explicitly calculated.
The explicit expressions for the two-point correlators can be found 
in section 5.
Section 6 is devoted to the comparison with the results of continuous
(Liouville) approach and to the conclusions.

\section{ Toda hierarchy for  the compactified 2D string}

\subsection{ Matrix model of 2D string and coupling to the windings}

    In this section we recall the main results of the paper \cite{KKK}
leading to the Toda hierarchy description for the compactified 2D
string theory. 

This theory   is represented by the Polyakov string action:
\be
S(x,g)={1\over 4\pi}\int d^2\sigma\sqrt{{\it det} g}
[ g^{ab}\p_a x\p_b x +\tmu],
\label{STR}
\ee
where $g_{ab}$ is a two dimensional fluctuating world sheet metric
and the only bosonic field $x(\sigma)$ is compactified on a circle of
a radius $R$: 
\be 
x(\sigma)\sim x(\sigma)+2\pi R .
\ee
 Such a field admits in general  Berezinski-Kosterlitz-Thouless
vortices (windings) in its configurations. If we denote by $x^{n_{\pm
1},n_{\pm 2},...}(\sigma)$ a configuration of the bosonic field
containing $n_{\pm 1}$ vortices of vorticity charge $\pm 1$, $n_{\pm 2}$
vortices of vorticity charge $\pm 2$, etc, the partition function of the
theory in this sector will be
\be
Z_{n_{\pm 1},n_{\pm 2},\cdots}=\int D g(\sigma)\int D x^{n_{\pm
1},n_{\pm 2},...}(\sigma) e^{-S(x,g)}.
\label{ZVOR}
\ee
Obviously, only the total vorticity $\sum_{k=-\infty}^{\infty} k n_{k}$
is conserved by topological reasons, but we will distinguish
configurations with different individual vortices as point-like
objects with vorticities $k$, to give a statistical-mechanical
meaning to the system.

Instead of fixing the numbers of vortices $n_k$ we can also introduce the
partition function with fixed fugacities $t_{\pm 1}, t_{\pm 2}, \dots$ 
of the vortices with given charges:
\be 
Z_{t_{\pm 1},t_{\pm 2},\cdots}=\sum_{n_{\pm 1},n_{\pm 2},\cdots}
t_{\pm 1}^{n_{\pm 1}} t_{\pm 2}^{n_{\pm 2}}\cdots Z_{n_{\pm 1},n_{\pm
2},\cdots}.
\label{GENF}
\ee

It is well known that in the case of flat 2D space
($g_{ab}=\delta_{ab}$) such a system can be described in a dual,
coulomb gas picture, in terms of a 2D field theory of one massless
scalar field $\tilde x(\sigma)$ (not necessarily compactified)
perturbed by vortex operators built out of this field. In the case of
fluctuating metric this scalar field gets coupled to the Liouville
field. The action (at least in the small coupling regime
$\phi\to\infty$) looks as
\be
 S={1\over4\pi}\int d^2 z\, [(\partial x)^2 +(\partial\phi)^2
 -4\hat \CR\phi+ \tmu\phi e^{-2\phi}
  + \sum_{n\ne 0}  \tln  e^{(  |n|R-2)\phi}  
e^{i n R \tilde x }],
\label{LVORT}
\ee
where $x=x_L+x_R, \ \ \tilde x=x_L-x_R$ ($x_L$ and $x_R$ are left
and right movers).
The total curvature is normalized as 
$\int d^2 z\, \sqrt{g}\hat \CR = 2\pi (2-2h)$, where $h$ is
the genus of the surface.
Note that we chose the central
charge $c=26$ and the Liouville ``charges'' of vortex operators to
make them marginal. The compactification radius $R$ remains a free
parameter of the theory.  If we choose the couplings as
\be
\tln= \frac{\tl}{2}(\delta_{n,1} +\delta_{n, -1}),
\label{TDEF}
\ee
the model will turn into the Sine-Gordon theory coupled to 2D
gravity:
\be
 S_{SGG}={1\over4\pi}\int d^2 z\, [(\partial x)^2 +(\partial\phi)^2
 -4\hat \CR\phi+ \tmu\phi e^{-2\phi}
  +\tl  e^{(R-2)\phi}  
\cos{  R \tilde x }]. 
\label{LSG}
\ee
  As its
counterparts on the flat space, this theory is supposed to be in the
same universality class as the compactified 2D string theory
(\ref{STR}).

In particular case of vanishing cosmological constant $\tmu$ and
$R=3/2$ (here $R_{KT}=2$)
the CFT (\ref{LSG}) was conjectured \cite{FZZ} to be dual to 
the coset CFT describing the string theory on 
the dilatonic black hole background \cite{Wit, DVV} 
(see \cite{KKK} for the details). 

The matrix model version of this theory is represented by the 
 partition function of the MQM on a ``time'' circle of the
radius $R$
\be
\label{measom}
\CZ_N[ \lambda,g]=
 \int [d\Omega]_{SU(N)}\  \exp\left( \sum_{n\in\Z} \lambda
_n \tr \Omega^n\right) \CZ_N(\Omega,g),
\ee
where
\be
\label{MMM}
 \CZ_N(\Omega,g)
= \int_{M(2\pi R)=\Omega^{\dagger} M(0)\Omega} {\CD}M(x) e^{ - \tr
\int ^{2\pi R} _0 dx \left[ \hf (\p_x M)^2  +V(M)\right] }.
\ee
We take the twisted boundary conditions for the hermitian matrix field
$M(x)$, where the twist matrix $\Omega$ can be chosen without lack of
generality as a Cartan element of the $SU(N)$ group
$\Omega={diag}(z_1,z_2,\cdots,z_N)$. The matrix potential is,
for example, $V(M) = \hf M^2 - {g\over 3\sqrt{N}}M^3$. 
The couplings $\lambda_n$ will play the same role here as $\tln$
in (\ref{LVORT}) although the exact relation between them depends
on the regularization procedure (the shape of the potential $V(M)$). 

Following the usual logic of the ``old'' matrix models, the Feynman
expansion of this model can be shown (see \cite{KKK} for the details)
to describe the compactified lattice scalar field $x_i$ living in the
vertices $i=1,2,3,\dots$ of $\phi^3$ type planar graphs. This field
can have vortex configurations: the vortices of charges $n=\pm 1, \pm
2, \dots$ occur on the faces of planar graphs and are weighted with
the factors $\lambda_n$. So the model (\ref{measom}) represents 
a lattice analogue
of the continuous partition function for the models
(\ref{STR}), (\ref{GENF}) or (\ref{LVORT}). 
The sum over planar graphs represents the functional
integral over two dimensional metrics of the world sheet and due to
the standard `t Hooft argument $1/N$ expansion goes over the powers
$1/N^{2-2h}$ where $h$ is a genus of the world sheet.

This model has a long history. In \cite{GRKLbis} some estimates of
the matrix approach were compared to the predictions of (\ref{LSG})
and appeared to be in full agreement. In \cite{BULKA} the model
(\ref{STR}) was formulated in terms of a twisted inverted matrix
oscillator (\ref{MMM}). The Toda hierarchy description of the model
(for the usual stable oscillatorial potential) was proposed in
\cite{HKK}. In the paper \cite{MOORE} the dual version of the matrix
model of \cite{GRKL,GRKLbis}, the Sine-Gordon model on random graphs,
was used to find (or rather correctly conjecture) the free energy of
(\ref{LVORT}) as a function of the cosmological constant $\tmu$,
Sine-Gordon coupling $\tl$ and radius $R$.

The twisted inverted matrix oscillator of \cite{BULKA} appeared to be
very useful for the identification of the free energy of the whole
theory (\ref{LVORT}) and of the corresponding matrix model (\ref{measom}) 
with the $\tau$-function of the Toda chain
hierarchy of integrable PDF's \cite{KKK}. 
To see how this description emerges
 we pass to the grand canonical partition function:
\be
 \CZ_\mu[\lambda,g] =\sum _{N=0}^{\infty} e^{-2\pi
R \mu N}\  \CZ_N[\lambda,g].
\label{DKPF}
\ee
The chemical potential $\mu$ happens to be the cosmological constant
similar to $\tilde\mu$ appearing in (\ref{LSG}) (see \cite{KLrev} and
\cite{BULKA} for the details).  Then following the double scaling
prescription for the $c=1$ matrix model \cite{DSL} corresponding to
the continuous limit for the lattice world sheets we should send in
(\ref{measom}) $N\to\infty$ and $g\to g_{crit}$ in such a way that at
the saddle point of the sum over $N$ in (\ref{DKPF})
$$
\mu={1\over 2\pi R}{\p\over \p N}\log \CZ_N[\lambda,g]
$$ 
remains fixed. This amounts to shifting $M\to M+{\sqrt{N}\over g}$  and
neglecting the cubic term in the potential $V(M)$, or rather treating
it as a cut-off wall at $M\to\infty$, which gives $V(M)\simeq -\hf
M^2$. Then the integral \cite{BULKA} can be calculated exactly, giving:
\be
\label{canpart}
\CZ_N[\lambda]={1 \over N!}\oint \prod_{k=1}^{N} 
\ {dz_k\over 2\pi i z_k} \ {e^{ 2 u(z_k)}
\over ( q^{1/2}-
q^{-1/2} ) } \ \prod_{ j \ne j' } ^{N} {z_j \ -\ z_{j'} \over q^{1/2}
z_j - q^{-1/2} z_{j'}}. \label{ZN}
\ee
where $u(z)=\hf\sum_n \lambda_n z^n$, $q=e^{2\pi iR}$ .

For the identification with $\tau$-function it is convenient to redefine
the vortex couplings as
\be 
t_n=\frac{\lambda_n}{q^{n/2}-q^{-n/2}}. 
\label{tn} 
\ee
Then plugging (\ref{canpart}) into (\ref{DKPF}) we recognize  in it the
$\tau$-function of Toda lattice hierarchy \cite{Hir, UT}
\be
\label{tauf}
\tau_l[t] = e^{-\sum _n nt_{n}t_{-n} } 
\sum_{N=0}^{\infty} (q^{ l}
\ e^{-2\pi R \mu } )^{N}  \CZ_N[t]=e^{-\sum _n nt_{n}t_{-n}}
\CZ_{\mu-il}[t],
\ee
where the charge $l$ giving an extra ``lattice'' dimension to the Toda
equations appears to be an imaginary integer shift of the cosmological
constant $\mu$.
 The couplings $t_n$ of vortices, 
together with $t_0=\mu$ turn out to be the ``times'' of commuting
flows of the hierarchy. Due to this the whole sector of
the theory describing the dynamics of winding modes of the model
(\ref{STR}) is completely solvable and the calculations of particular
physical quantities,  such as the correlators of vortex operators, are
greatly simplified. In particular, the $\tau$ function (\ref{tauf}) represents the
generating function for such correlators in the theory (\ref{LSG}):
\be
\K_{i_1 \cdots i_n}=\left.
\frac{\partial^{n}}{\partial \lambda_{i_1} \cdots \partial
\lambda_{i_n}} \log\tau_0\right|_{\lambda_{\pm 2}=\lambda_{\pm 3}=\cdots =0}.
\label{correlat}
\ee
with $\mu$ and $\lambda_{\pm 1}$ fixed.

Using the Hirota equations of Toda hierarchy we will calculate in this
paper the one-point and two-point functions of vortex operators of
arbitrary vorticities in the spherical approximation.

\subsection{ Hirota equations}

One can show \cite{JM} that the ensemble of the $\tau$-functions of the Toda
hierarchy with different charges satisfies 
a set of bilinear partial differential 
equations known as Hirota equations which are written as follows
(the derivatives are taken with respect to $t_n$ rather than $\lambda_n$)
\begin{eqnarray}
&\sum\limits_{j=0}^{\infty}p_{j+i}(-2y_+)p_j(\tilde D_+)
\exp \left( \sum\limits_{k\ne 0}y_kD_k\right) 
\tau_{i+l+1}[t]\cdot\tau_l[t] =
& \nonumber \\
&\sum\limits_{j=0}^{\infty}p_{j-i}(-2y_-)p_j(\tilde D_-)
\exp \left( \sum\limits_{k\ne 0}y_kD_k\right) 
\tau_{i+l}[t]\cdot\tau_{l+1}[t],&
\label{toda}
\end{eqnarray}
where
\beq
y_{\pm} &=& (y_{\pm 1}, y_{\pm 2},y_{\pm 3},\dots), \\
\tilde D_{\pm} &=& (D_{\pm 1}, D_{\pm 2}/2,D_{\pm 3}/3,\dots)
\eeq
represent the Hirota's bilinear operators
\be
D_n f[t]\cdot g[t]=\left. \frac{\partial}{\partial
x}f(t_n+x)g(t_n-x)\right|_{x=0},
\ee
and $p_j$ are Schur polynomials defined by
\be
\sum\limits_{k=0}^{\infty}p_k[t]x^k=
\exp\left(\sum\limits_{k=1}^{\infty}t_n x^n\right). 
\ee
Since the $\tau$-function of the Toda hierarchy is related to the free
energy by  
\be \tau_s[\mu,t]=\exp\left(F(\mu-is)\right),\label{tau-f} \ee
the Hirota equations lead to a triangular system of nonlinear 
difference-differential equations for the free energy of the model
(\ref{DKPF})-(\ref{ZN}). 
Due to (\ref{correlat}) it can be actually thought of as a system
 of equations for the correlators which we are interested in.

\subsection{ Scaling of winding operators} 

It turns out that there is a scale in the model which corresponds 
to the scale given by the string coupling constant in the string theory.
Since $g_{s}\sim e^{<\phi>}$ it can be associated with 
the Liouville field $\phi$
or the cosmological constant $\tmu$ coupled with it.
All couplings $\tln$ have definite dimensions with respect to this scale.
In the Toda hierarchy the counterparts of $\tmu$ and $\tln$
are $\mu$ and $t_n$ correspondingly.
As one can immediately see from (\ref{LVORT}) by shifting the zero
mode of the Liouville field, the corresponding dimensions 
of the couplings
$t_n$ with respect to the rescaling of the cosmological constant
$\mu$ are
\be \Delta[t_n]=1-\frac{R|n|}{2}.  
\label{dim} 
\ee

This scaling leads to an expansion for the free energy which can
be interpreted as its genus expansion
\beq 
F&=&\sum\limits_{h=0}^{\infty}F_h, \label{fe} \\
F_0&=&\xi^{-2}f_0(\w ;s)+\frac{R}{2}\mu^2\log\xi, \label{F0} \\
F_h&=&\xi^{2h-2}f_h(\w ;s), \qquad h>0, \eeq 
where the quantities $f_h$ are the functions of the dimensionless parameters 
\beq & s=(s_2,s_3,
\dots), \qquad s_n=i\left(-\frac{t_{-1}}{t_{1}}\right)^{n/2}
\xi^{\Delta[t_n]}t_n,    & \label{sss}\\ 
& \w=\mu\xi, \quad {\rm with} \quad
\xi=(\lambda\sqrt{R-1})^{-\frac{2}{2-R}}=
(t_1t_{-1}(R-1))^{-\frac{1}{2-R}}. & 
\label{yxi} 
\eeq
Remarkably, the Toda equations (\ref{toda}) are  compatible with
this scaling.  

The spherical limit of the string partition function is described by
the dispersionless Toda hierarchy which can be obtained taking $\xi
\longrightarrow 0$.  In this limit the explicit expression for the
partition function with vanishing $s_n$ has been conjectured in
\cite{MOORE} and proven in \cite{KKK} . The result formulated in
terms of $X_0(\w)=\partial^2_{\w}f_0(\w)$ reads
\be
w=e^{-\frac{1}{R}X_0}-e^{-\frac{R-1}{R}X_0} 
\label{yyy}
\ee
or, integrating $X_0$ twice over $\w$
\be
F_0=
 \frac12 \mu^2\left( R\log \xi+
 X_0 \right)   
+\xi^{-2}\left({3\over 4} {R\over R-1} 
e^{-2{R-1\over R} X_0}+{3\over 4}R e^{- {2\over R}X_0}-
{R^2-R+1\over R-1} e^{- X_0} \right).    
\label{FREN}
\ee
This result will serve as an input for the following calculation of
the correlators in the spherical limit.

\section{Two-point correlators from the Toda hierarchy}

\subsection{Equation for the generating function of two-point correlators}

To find equations for the two-point correlators in the spherical
limit let us take in (\ref{toda}) $i=0$ and the coefficient in front
of $y_{n}y_{m},\ n,m>0$.  Then we obtain the following equation 
\beq
&\left[ 4p_{n+m}(\tD_+)-2p_n(\tD_+)D_{m} -2p_m(\tD_+)D_{n}
+D_nD_m\right]\tau_{s+1}\cdot\tau_{s}= D_nD_m\tau_{s}\cdot\tau_{s+1},
& \nonumber \\ &\left[ 2p_{n+m}(\tD_+)-p_n(\tD_+)D_{m}
-p_m(\tD_+)D_{n} \right]\tau_{s+1}\cdot\tau_{s}=0. & 
\label{eq1}
\eeq
This equation is valid also at $m=0$ or $n=0$ if we set $D_0\equiv 1$.
Such an equation can be obtained from the coefficient in front of
$y_n$ in (\ref{toda}).

If we multiply the equation (\ref{eq1})
by $x^ny^m$ and sum over all $n$ and $m$ we obtain
\beq 
&\left[ 2\sum\limits_{n,m=0}^{\infty} x^ny^m p_{n+m}(\tD_+)-
\sum\limits_{m=0}^{\infty} y^m D_{m} 
\exp\left\{ \sum\limits_{n=1}^{\infty}x^n \tD_n\right\}  \right. & \nonumber \\
& \left. -
\sum\limits_{n=0}^{\infty} x^nD_{n} 
\exp\left\{ \sum\limits_{m=1}^{\infty}y^m \tD_m\right\}\right]
\tau_{s+1}\cdot\tau_{s}=0.&
\label{eq2}
\eeq  
Furthermore, it is easy to see that the first term can be 
rewritten as follows:
\be
\frac{2}{x-y}\left[ x e^{\sum\limits_{k=1}^{\infty}x^k \tD_k}
- y e^{\sum\limits_{k=1}^{\infty}y^k \tD_k} \right]\tau_{s+1}\cdot\tau_{s}. 
\label{t1}
\ee

 It is easy to check the following remarkable formula
\be f(D)e^F\cdot e^G=e^Fe^G f\left(D+(\partial F\cdot 1-
1\cdot\partial G)\right), \ee
which allows to express $\tau$-functions in the equation (\ref{eq2}), 
(\ref{t1}) through the derivatives of the free energy:
\beq
& \frac{2}{x-y}\left( x e^{\sum\limits_{k=1}^{\infty}x^k (\tD_k+\tX_k)}
- y e^{\sum\limits_{k=1}^{\infty}y^k (\tD_k+\tX_k)} \right)= &
\nonumber \\
&  \left( 1+\sum\limits_{m=1}^{\infty} y^m (D_{m}+X_m)\right) 
e^{ \sum\limits_{k=1}^{\infty}x^k (\tD_k+\tX_k)} +
\left( 1+\sum\limits_{n=1}^{\infty} x^n(D_{n}+X_n) \right)
e^{ \sum\limits_{k=1}^{\infty}y^k (\tD_k+\tX_k)}, &
 \label{eq3}
\eeq
where we have introduced the notation $\tX_k=\frac{1}{|k|}X_k$ and
\footnote{We can shift the index $s$ of the free energy 
$F_s=\log \tau_s$ since due to (\ref{tau-f}) the resulting equation
depends only on the difference of the indices of two $\tau$-functions.}
\be 
X_k=\partial_k F_{\frac12}\cdot 1 -
1\cdot \partial_k F_{-\frac12}, \qquad k\ne 0. \label{gX0n}
\ee 
The $1$'s in the right hand side of (\ref{eq3})
arise from the definition of $D_0$.
 
We show in Appendix A that in the dispersionless limit only the second 
derivatives of the free energy survive.
This means that we can put the second and higher derivatives 
of $X_k$ to zero: $D^n X_k=0$, $n>1$.
Besides, using (\ref{tau-f}) we obtain in this limit
\beq
\tX_n&=& -i\frac{1}{|n|}\frac{\partial^2 }{\partial t_n
\partial \mu} F_0 := \tX_{0,n},  \label{X0n} \\
\tD_n \tX_m &  =& 2\frac{1}{|n||m|}\frac{\partial^2}
{\partial t_n \partial t_m} F_0
:= \tX_{n,m} .    \label{Xnm}
\eeq
Thus the fields $\tX_{n,m}$ correspond (up to a numerical factor) to
the two-point correlators, whereas $\tX_{0,n}$ give the one-point ones.
We also define the generating function of all $\tX$'s
\be F(x,y)=\sum\limits_{n,m=0}^{\infty} x^n y^m \tX_{n,m}, \label{genf} \ee
where we have taken $\tX_{0,0}=0$.

With these definitions, trivial manipulations which can be found in
Appendix B, lead to the following equation for the generating function
\be
\frac{x+y}{x-y}\left( 
e^{\frac12 F(x,x)} - e^{\frac12 F(y,y)} \right) 
= x\partial_x F(x,y) e^{\frac12 F(y,y)} +
y\partial_y F(x,y) e^{\frac12 F(x,x)}.  \label{eqMain}
\ee

\subsection{Solution in terms of one-point correlators}

It turns out that the differential equation (\ref{eqMain}) is solvable.
First of all, change the variables to
\beq
d\eta&=&e^{\frac12 F(x,x)}\frac{dx}{x}+ e^{\frac12 F(y,y)}\frac{dy}{y} ,
\nonumber \\
d\zeta&=&e^{\frac12 F(x,x)}\frac{dx}{x}- e^{\frac12 F(y,y)}\frac{dy}{y} .
\label{newvar}
\eeq
It is easy to see that 
\be 
\zeta=\int\limits_y^x \frac{dz}{z}e^{\frac12 F(z,z)}=
\log \frac{x}{y}+h(x)-h(y),
\label{zeta}
\ee
where
\be h(x)=\int\limits_0^x\frac{dz}{z}\left(e^{\frac12 F(z,z)}-1\right) 
\label{hx}
\ee
is a regular function at $x=0$.

In terms of the new variables the equation reads
\be
\partial_{\eta} F(x,y)= \frac12\frac{x+y}{x-y}\left( 
e^{-\frac12 F(y,y)} - e^{-\frac12 F(x,x)} \right) 
=\partial_{\eta} \log \frac{xy}{(x-y)^2}.
\ee
Thus
\be F(x,y)= \log \frac{xy}{(x-y)^2}+ g(\zeta), \label{Fsol} \ee
where the function $g(\zeta)$ is fixed by the requirement for $F(x,y)$
to be analitycal in both arguments. 

The analitycity condition together with (\ref{zeta}) leads to the following
requirements on $g(\zeta)$:
i)~$g(\zeta) \mathop{\sim}\limits_{\zeta \to \pm\infty} \pm\zeta$; 
ii) $g(\zeta) \mathop{\sim}\limits_{\zeta \to 0} \log\zeta^2$.
There is only one function satisfying both of them,
which is given by
\be
g(\zeta)=\log \left( e^{\zeta}+e^{-\zeta}-2\right)=
\log\left( 4{\rm sh}^2\left(
\frac{\zeta}{2}\right)\right).  \label{gsol}
\ee
Taking together (\ref{Fsol}), (\ref{gsol}) and (\ref{zeta}) we obtain
the solution for the generating function of the two-point correlators
in terms of a still unknown regular function $h(x)$
\be 
F(x,y)=\log\left[ \frac{4xy}{(x-y)^2}{\rm sh}^2\left(\frac{1}{2}
(h(x)-h(y)+\log\frac{x}{y})\right)\right]. \label{Fexp}
\ee
From (\ref{Fexp}) it is easy to see that $h(x)$ is nothing else than 
the generating function for the one-point correlators since $h(x)=F(x,0)$.

\subsection{Two-point correlators with vorticities of opposite signs}

Up to now we considered the two-point correlators 
for vorticities of positive sign only. 
Now we generalize the results for other cases.

First of all, the case of both negative signs can be directly
obtained from the previous one. The corresponding equation which is derived
from the coefficient in front of $y_{-n}y_{-m}$ in
the Hirota equation looks as (\ref{eq1})
\be
\left[ 2p_{n+m}(\tD_-)-p_n(\tD_-)D_{-m}
-p_m(\tD_-)D_{-n} \right]\tau_{s}\cdot\tau_{s+1}=0. 
 \label{eqneg1}\ee
The only difference is that $\tau_{s}$ and $\tau_{s+1}$ are exchanged.
Due to this there is an additional minus in the definition of $\tX_{0,-n}$.
So we should distinguish two types of correlators
\beq & \tX^{\pm}_{0,m} := \mp i
\frac{1}{|n|}\frac{\partial^2 }{\partial t_n \partial \mu} F_0,
 \quad n\ne 0, &  \label{X1pm}  \\
& \tX^{\pm}_{n,m} := 2\frac{1}{|n||m|}\frac{\partial^2 }{\partial t_n
\partial t_m} F_0, \quad n,m\ne 0.
  &  \label{X2pm} 
\eeq
With this definition, the generating functions of the correlators for 
positive and negative vorticities are 
\be F^{\pm}(x,y)=\sum\limits_{n,m=0}^{\infty} x^n y^m 
\tX^{\pm}_{\pm n,\pm m},
\label{genf+} \ee
Despite of this minus all equations for the case of negative vorticities 
being written in terms of $\tX^-_{n,m}$ are the same as for the case
of positive ones.
As a result $F^-(x,y)$ is given by the same solution (\ref{Fexp})
as $F^+(x,y)$. The only difference between them is the generating
function of the one-point correlators $h$ which should be replaced in
(\ref{Fexp}) by $h^{\pm}$ correspondingly.

To find the correlators with vorticities of opposite signs, take the
coefficient in the Hirota equation in front of $y_{n}y_{-m}$ and set $i=1$.
 Then we obtain
the following equation
\beq  
& -2p_n(\tD_+)D_{-m}\tau_{s+2}\cdot\tau_{s}=
\left[ -2 p_m(\tD_-)D_{n} +\tD_{-1}D_{-m}D_n\right] 
\tau_{s+1}\cdot\tau_{s+1}, & \nonumber \\
  & p_n(\tD_+)D_{-m}\tau_{s+2}\cdot\tau_{s}=
p_m(\tD_-)D_{n} \tau_{s+1}\cdot\tau_{s+1}. &
 \label{eqd5}
 \eeq
The last term in the first line vanishes since it always gives rise to 
$D_k (F_{s+1}\cdot 1 - 1\cdot F_{s+1})=0$. In general, 
the one-point correlators appearing from $\tau_{s+k}\cdot \tau_s$ are
always supplied with the coefficient $k$ what follows from (\ref{tau-f}).
Due to this in the right hand side
there are no one-point correlators whereas in the left hand side
they enter with the coefficient 2.
This equation is valid also for $m=0$ if we take $D_0\equiv -2$
instead of the previous choice.
However, for $n=0$ it is never fulfilled. 

The arguments similar to those of Section 1 lead to 
the following equation for the generating function
\be
A\left[ y\partial_y (G(x,y)-2h^-(y))-2\right] e^{\frac12 F^+(x,x)+h^+(x)} =
\frac{1}{y}\partial_x G(x,y) e^{\frac12 F^-(y,y)-h^-(y)},  \label{eqd2}
\ee
where we have introduced
\beq 
& A=\exp\left( -\partial^2_{\mu} F_0\right)=
\xi^{-R}e^{-\partial^2_{\w} f_0} ,  & \label{constA}  \\
& G(x,y)=\sum\limits_{n,m=1}^{\infty} x^n y^m \tX_{n,-m}. &
\label{genfd} 
\eeq
Using (\ref{hx}) we can rewrite 
(\ref{eqd2}) as follows
\be \left( \left[ A \partial_x(xe^{h^+(x)})\right]^{-1}\partial_x +
\left[\partial_y(\frac{1}{y}e^{-h^-(y)})\right]^{-1} 
\partial_y \right) G(x,y) = 
 -2ye^{h^-(y)} .  \label{eqd3}
\ee
It is clear that the solution of this equation looks as
\be  G(x,y)=2\log \left( y e^{h^-(y)} \right) +
f(\frac{1}{y}e^{-h^-(y)}-Axe^{h^+(x)}).  \ee
The unknown function $f$ is determined by the analitycity of $G(x,y)$.
Requiring cancelation of the logarithmic singularity at $y=0$
we obtain that $f(z)=2\log z$,
As a result we find the generating function
\be 
G(x,y)=
2\log\left(1-Axy e^{h^+(x)+h^-(y)}\right).
\label{Dsolution} 
\ee

Let us note that the equations (\ref{Fexp}) and (\ref{Dsolution})
resemble the equations for two-point correlators of the two-matrix
model found in \cite{DKK}.

\section{One-point correlators}

\subsection{Equation for the generating function}

Now we should work out an equation for the one-point correlators.
We can always put  
$t_1=-t_{-1}$ (corresponding to 
$\lambda_1=\lambda_{-1}$, see (\ref{tn})).
From the equations written below for positive vorticities it is
clear that in this case
$h^{\pm}(x)$ coincide with each other. For this reason we omit
in the following the inessential sign label in $h(x)$.
The case of general $t$'s can be obtained from the previous one by
the substitution
$h^{\pm}(x)=h\left((-t_{-1}/t_{1})^{\pm 1/2}x\right)$. 
It can be easily seen from (\ref{sss})
and is due to the total vorticity conservation.
Accordingly, 
\beq
F^{\pm}(x,y)&=&F\left( (-t_{-1}/t_{1})^{\pm 1/2}x,
(-t_{-1}/t_{1})^{\pm 1/2}y\right), \\
G(x,y)&=& G_{t_1=-t_{-1}}\left( (-t_{-1}/t_{1})^{1/2}x,
(-t_{-1}/t_{1})^{-1/2}y\right).
\eeq

The equation we are looking for 
arises from two facts:  i) we know the dependence of
the free energy on $t_{\pm 1}$ and $\mu$ (\ref{yyy}),
(\ref{FREN}); ii) the system of 
equations for the correlators is triangular. Due to this 
we can express $\tX_{\pm 1,n}$ through $\tX_{0,n}$ by a linear 
integral-differential operator. Namely, in terms of the generating functions
we can write 
\beq 
\partial_y F^+(x,0)&=&\hat K^{(+)} h(x)+\tX^+_{1,0}, \label{eqh+}\\
\partial_y G(x,0)&=&\hat K^{(-)} h(x), \label{eqh+-}
\eeq
where the operators $\hat K^{(\pm)}$ are to be found.
The last term in the first line
is explicitly added since in our notations $F^+(0,0)=\tX^+_{0,0}=0$.
Moreover, we can rewrite these equations in terms of the generating
function for the one-point correlators only. 
Indeed, from (\ref{Fexp}) and (\ref{Dsolution}) we obtain
\beq
\hat K^{(+)} h(x)&=&\frac{2}{x}\left( 1- e^{-h(x)}\right) -2\tX^+_{0,1},
\label{p1+}  \\
\hat K^{(-)} h(x)&=&-2Axe^{h(x)}.
\label{p1-} 
\eeq   

To find the operators $\hat K^{(\pm)}$ we compare the derivatives of the free
energy with respect to $\mu$ and $t_{\pm 1}$. 
From (\ref{F0}), (\ref{sss}) and (\ref{yxi}) we obtain for $|n|>1$
\beq
\left. \frac{\partial}{\partial \mu}\left( \frac{\partial}{\partial t_n} F_0
\right)\right|_{s_k=0}&=&
i\left. \xi^{-\frac{|n|R}{2}} 
\frac{\partial}{\partial \w}\left(\frac{\partial}{\partial s_n}
f_0(\w;s) \right)\right|_{s_k=0}, \label{derivmu} \\
 \left. \frac{\partial}{\partial t_{\pm 1}}
\left( \frac{\partial}{\partial t_n} F_0
\right)\right|_{s_k=0}&=& 
\left. 
\mp\frac{\xi^{-\frac{(|n|+1)R}{2}}}{2-R}
\left(\w\frac{\partial}{\partial \w}-\left( 1+\frac{R|n|}{2}
\pm \frac{(R-2)n}{2} \right)
\right)\left(\frac{\partial}{\partial s_n}
f_0(y;s) \right)\right|_{s_k=0}.  \label{derivt1}
\eeq
These relations together with (\ref{X0n}) and (\ref{Xnm}) imply that the 
operators are given by
\beq
\hat K^{(+)}&=&-a
\left[\w-(1+(R-1)x\frac{\partial}{\partial x} )
\int\limits^{w} d\w\right], 
\label{opK+} \\
\hat K^{(-)}&=&a
\left[\w-(1+x\frac{\partial}{\partial x} )
\int\limits^{\w} d\w\right], 
\label{opK-}
\eeq
where 
$a =2\frac{\sqrt{R-1}}{2-R}\xi^{-\frac{R}{2}}$.
This expression is also valid for the case $n=\pm 1$ where 
the derivatives of the free energy contain additional
terms appearing from the logarithmic term in (\ref{F0}).

Inserting (\ref{opK+}) and (\ref{opK-}) into 
(\ref{p1+}) and (\ref{p1-}) and taking the derivative with
respect to $\w$ we obtain two differential equations for the generating
function of the one-point correlators
\beq
\left[ -a
\left( (R-1)x\frac{\partial}{\partial x} - \w
\frac{\partial}{\partial \w} \right) +\frac{2}{x}e^{-h(x;\w)}
\frac{\partial}{\partial \w}\right] h(x;\w)
&=&2\frac{\partial}{\partial \w}\tX^+_{0,1}, 
\label{p2+}  \\
\left[ a
\left( x\frac{\partial}{\partial x} - \w 
\frac{\partial}{\partial \w} \right) -2Ax e^{h(x;\w)}
\frac{\partial}{\partial \w}\right] h(x;\w)
&=&2x e^{h(x;\w)}\frac{\partial}{\partial \w}A. 
\label{p2-}
\eeq

\subsection{Solution}

Let us take advantage of having two equations for one quantity.
Using (\ref{p2+}) and (\ref{p2-}) we can exclude the derivative 
with respect to $x$. The resulting equation is
\be
\left( a(2-R)\w+2 \frac{e^{-h}}{x}-
2(R-1)Axe^{h}\right)\partial_{\w} h=
2\partial_{\w}\tX^+_{0,1}
+2(R-1)\partial_{\w}A xe^{h}.
\label{p3}
\ee

From (\ref{X1pm}), (\ref{F0}) and (\ref{FREN}) one can obtain
\be
\tX^+_{0,1}=
\left. -i \frac{\partial^2}{\partial \mu \partial
t_1} F_0 \right|_{s_k=0}=\frac{\sqrt{R-1}}{2-R}\xi^{-R/2}
\left( (\w \partial_{\w}-1)\partial_{\w}f_0(\w)+R\w\right)
=
\frac{R}{\sqrt{R-1}}\xi^{-R/2}e^{-\frac{R-1}{R}X_0}. 
\label{X01}
\ee
Noting that the equation (\ref{p3}) is homogeneous in the derivative
$\partial_{\w}$ we can change it by $\partial_{X_0}$.
Then after the substitution of the definitions of all entries
the common multiplier
$$2\sqrt{R-1}\xi^{-R/2} \left(
e^{-\frac{R-1}{R}X_0}+
\sqrt{R-1}\xi^{-R/2}
 x e^{h-X_0}\right)$$
can be canceled. 
As a result we obtain the following simple equation
\be
\left(1 -\frac{\xi^{R/2}}{x\sqrt{R-1}}
e^{\frac{R-1}{R}X_0}e^{-h}\right)\partial_{X_0}h=1.
\label{p4}
\ee
This equation can be trivially integrated giving
after the proper choice of the integration constant
\be
e^{\frac{1}{R}h}-ze^{h}=1, \label{hsol}
\ee
where $z=\frac{\xi^{-R/2}}{\sqrt{R-1}}
xe^{-\frac{R-1}{R}X_0}$.
As it is easy to check this solution satisfies both equations
(\ref{p2+}) and (\ref{p2-}).

\subsection{One-point correlators of fixed vorticities}

Explicit expressions for the one-point correlators are given by
the coefficients of the expansion in $x$ by the generating function $h(x)$ 
satisfying equation (\ref{hsol}).
To find them we can use
equation 5.2.13.30 in \cite{BPM}
\beq
 \frac{1}{b}s^b&=&\sum\limits_{n=0}^{\infty}
\frac{z^n}{n!}\frac{\Gamma(na+b)}{\Gamma(n(a-1)+b+1)}, \\
s&=&1+zs^a.
\eeq
Taking $s=e^{\frac{1}{R}h}$, $a=R$ we obtain for $b=kR$
\be
e^{kh}=kR\sum\limits_{n=0}^{\infty}
\frac{z^n}{n!}\frac{\Gamma((n+k)R)}{\Gamma((n+k)R-n+1)}. \label{exp1p}
\ee
Also the limit $b\rightarrow 0$ gives
\be
h=R\sum\limits_{n=1}^{\infty}
\frac{z^n}{n!}\frac{\Gamma(nR)}{\Gamma(n(R-1)+1)}.
\ee
This means that the derivative with respect to $\mu$ of
the one-point correlator is
\be
\frac{\partial^2}{\partial \mu \partial t_n}F=
\frac{i\Gamma(nR+1)}{n!\Gamma(n(R-1)+1)}
\frac{\xi^{-nR/2}}{(R-1)^{n/2}} e^{-n\frac{R-1}{R}X_0}.
\label{der1point} 
\ee
After integration over $\mu$ and taking into account the relation
(\ref{tn}) we obtain the following one-point correlators of operators
of vorticity $n$ in the spherical limit
\be
\K_n=
\frac{1}{2\sin \pi n R}
\frac{\Gamma(nR+1)}{n!\Gamma(n(R-1)+1)}
\frac{\xi^{-\frac{nR+2}{2}}}{(R-1)^{n/2}} 
\left(\frac{e^{-\frac{n(R-1)+1}{R}X_0}}{n(R-1)+1}
-\frac{e^{-(n+1)\frac{R-1}{R}X_0}}{n+1}\right).
\label{onepoint} 
\ee
(Here the lower limit of the integration is chosen to be $X_0=+\infty$.
Only this choice reproduces $\partial_{\lambda_1}F_0$ from (\ref{FREN}).) 
In particular, in the limit $\mu \rightarrow 0$ ($X_0 \rightarrow 0$)
corresponding to string theory on the black hole background  
(in fact, only the point $R=3/2$ is supposed to describe the 
conventional black hole) we find
\be
\K_n=
\frac{1}{2\sin \pi n R}
\frac{\xi^{-\frac{nR+2}{2}}}{(R-1)^{n/2}} 
\frac{(2-R)n\Gamma(nR+1)}{(n+1)!\Gamma(n(R-1)+2)}.
\label{bh1point} 
\ee

\section{Results for two-point correlators}

\subsection{Correlators with vorticities of opposite signs}

These correlators are given by the expansion coefficients of
the generating function (\ref{Dsolution}). This function can be 
easily expanded in $e^h$ for which the series (\ref{exp1p}) can be used.
This leads to the following expansion
\beq
D(x,y)&=&-2\sum\limits_{k=1}^{\infty}\frac{B^k}{k}z^k(x)z^k(y)
\sum\limits_{n=0}^{\infty}\sum\limits_{m=0}^{\infty}C_n^kC_m^k
z^n(x)z^m(y)
\nonumber \\
&=& 
-2\sum\limits_{n,m=1}^{\infty}z^n(x)z^m(y)
\sum\limits_{k=1}^{{\rm min}(n,m)}\frac{B^k}{k}C^k_{n-k}C^k_{m-k},
\eeq
where
\beq
B&=&A(R-1)\xi^Re^{2\frac{R-1}{R}X_0}=(R-1)e^{\frac{R-2}{R}X_0}, \\
C_n^k&=&\frac{kR}{n!}\frac{\Gamma((n+k)R)}{\Gamma(n(R-1)+kR+1)}.
\eeq
The two-point correlators appear to be
\beq
\K_{n,-m}&=& -
\frac{\Gamma(nR+1)}{2\sin \pi n R}\frac{\Gamma(mR+1)}{2\sin \pi m R}
\frac{\xi^{-(n+m)R/2}}{(R-1)^{(n+m)/2}} e^{-(n+m)\frac{R-1}{R}X_0}
\times \nonumber \\
&\times &
\sum\limits_{k=1}^{{\rm min}(n,m)}\frac{k(R-1)^k
e^{k\frac{R-1}{R}X_0}}
{(n-k)!(m-k)!\Gamma(n(R-1)+k+1)\Gamma(m(R-1)+k+1)}.
\label{pm2pc}
\eeq
To get the result in the black hole limit it is enough to set $X_0=0$
in the above expression.

It is worth to give also the result for the ratios of correlators
which do not depend on possible leg-factors --- wave 
function renormalisations of the operators 
(see for ex. \cite{KLrev}):
\be
\frac{\K_{n,-m}}{\K_n\K_{-m}}= -\frac{\xi^2}{(2-R)^2 nm}
\sum\limits_{k=1}^{{\rm min}(n,m)}\frac{k(R-1)^k
(n+1)!(m+1)!\Gamma(n(R-1)+2)\Gamma(m(R-1)+2)}
{(n-k)!(m-k)!\Gamma(n(R-1)+k+1)\Gamma(m(R-1)+k+1)}.
\label{pm2pcn}
\ee
 Since they can in principle be compared
with CFT calculations in the black hole limit 
we restrict ourselves only to this particular case.

\subsection{Correlators with vorticities of same signs}

To investigate this case we should
expand (\ref{Fexp}) in $x$ and $y$.
The expansion in the first argument gives for $n>0$
\beq 
\partial_x^n F(0,y)&=&\frac{2}{y^n}\left(\frac{1}{n}-
\sum\limits_{k=1}^n
\frac{2}{k}z^{n-k}(y)e^{-k h(y)}C^{k}_{n-k}\right)-\tX_{0,n}
\nonumber \\
&=& -2\frac{\xi^{-nR/2}}{(R-1)^{n/2}}e^{-n\frac{R-1}{R}X_0}
\sum\limits_{m=0}^{\infty}z^m(y)\sum\limits_{k=1}^n\frac{1}{k}
C_{n-k}^{k} C_{m+k}^{-k}-\tX_{0,n}
\nonumber \\
&+& \frac{2}{y^n}\left( \frac{1}{n}-
\sum\limits_{m=1}^n z^{n-m}(y)\sum\limits_{k=m}^n\frac{1}{k}
C_{n-k}^{k} C_{k-m}^{-k}\right). \label{ppexp}
\eeq
The last term in the right hand side
is singular and we know from the analitycity of $F(x,y)$
that it should vanish. The remaining part gives the values of the 
two-point correlators   
\beq
\K_{n,m}&=& -
\frac{\Gamma(nR+1)}{2\sin \pi n R}\frac{\Gamma(mR+1)}{2\sin \pi m R}
\frac{\xi^{-(n+m)R/2}}{(R-1)^{(n+m)/2}} e^{-(n+m)\frac{R-1}{R}X_0}
\times \nonumber \\
&\times &
\sum\limits_{k=1}^{n}\frac{k}
{(n-k)!(m+k)!\Gamma(n(R-1)+k+1)\Gamma(m(R-1)-k+1)}.
\label{pp2pc}
\eeq
As in the previous case the black hole limit of (\ref{pp2pc})
is obtained vanishing $X_0$. The normalized correlators are
\be
\frac{\K_{n,m}}{\K_n\K_{m}}= -\frac{\xi^2}{(2-R)^2 nm}
\sum\limits_{k=1}^{n}\frac{k
(n+1)!(m+1)!\Gamma(n(R-1)+2)\Gamma(m(R-1)+2)}
{(n-k)!(m+k)!\Gamma(n(R-1)+k+1)\Gamma(m(R-1)-k+1)}.
\label{pp2pcn}
\ee

It is clear from the definition that (\ref{pp2pc})
as well as (\ref{pp2pcn}) should be symmetric in $m$ and $n$.
However, our results do not possess this symmetry explicitly.
The reason is that the expansion of (\ref{Fexp}) 
has been done in a nonsymmetric way.
It may be possible to symmetrize them
by means of identities between $\Gamma$-functions like
\be
\sum\limits_{k=m}^n\frac{1}{k}
C_{n-k}^{k} C_{k-m}^{-k}=0, \qquad 1\le m < n.
\ee
This identity follows from vanishing of the last term in (\ref{ppexp}).

\section{Discussion}

The main result of this paper is the calculation of
one- and two-point vorticity (winding) correlators
given by the generating functions (\ref{Fexp}), (\ref{Dsolution}) and
(\ref{hsol}).  
What kind of physics do they describe? 

We can look at our model from two points of view: as a
statistical-mechanical system describing a gas of vortices
and as a string theory.

\subsubsection*{Statistical mechanical picture}

 In the statistical mechanical picture  we interpret the
planar graphs of the MQM (\ref{MMM}) as random dynamical lattices
populated by Berezinski-Kosterlitz-Thouless (BKZ) vortices.

Our one point correlator $\K_n$ (\ref{onepoint}) should be in
principle proportional to the probability to find in the system a
vortex of a given vorticity $n$. The probability should be a positive
quantity. However, we see that the explicit expressions
(\ref{onepoint})-(\ref{bh1point}) contain the sign-changing factors
${1\over \sin\pi R n}$. Their origin is not completely clear to us but
we see two possible explanations for them:

i) They appeared as a result of wave function renormalisation in the
continuous (double scaling) limit and should be absorbed into the
``leg-factors''. The rest of the one-point correlator is strictly
positive in the interval $1<R<2$ which we pretend to
describe. This point of view is supported by the fact that these
factors appear as the same wave-function renormalisations also in the
two point correlators (\ref{pm2pc}),(\ref{pp2pc}). Since their origin
is completely due to the change of variables (\ref{tn}) these factors
appear to be a general feature for all multipoint correlators.

ii) The sign-changing might be a consequence of the fact that our
approach based on the Hirota equations (describing directly the double
scaling limit) does not keep track of non-universal terms due to the
ultraviolet (lattice) cut-off in the system. Such UV divergences are
well known in the Coulomb gas description of the BKT system: they
usually correspond to the proper energy of a vortex. In that case the
correlators we calculated above are only the universal parts of the
full correlators containing also big positive non-universal
contributions. These universal parts cannot be interpreted as full
probabilities and have no reason to be positively defined.

To prove or reject these explanations we have to see what happens with
the individual vortices in the process of taking the double scaling
limit, from planar graphs picture to the inverted oscillator
description. It is not an easy, although not hopeless, task which we
leave for the future.

\subsubsection*{String theory picture}

 The MQM describes also the two-dimensional bosonic string theory in
compact imaginary time (i.e. at finite temperature). Due to the FZZ
\cite{FZZ} conjecture, at the vanishing cosmological constant it can
be also interpreted as a string theory in the Euclidean 2D black hole
background (at least at $R=3/2$) \cite{KKK}.  Our one-point
correlators should contain information about the amplitudes of emission
of winding modes by the black hole, whereas the two-point correlators
describe the S-matrix of scattering of the winding states from the tip
of the cigar (or from the Sine-Liouville wall).  However, the exact
correspondence is lacking due to the absence of the proper
normalization of the operators.  In \cite{FZZ} and \cite{FB} some two-
and three-point correlators were computed in the CFT approach to this
theory. It would be interesting to compare their results with the
correlators calculated in this paper from the MQM approach. However
there are immediate obstacles to this comparison.

First of all, these authors do not give any results for the one point
functions of vortices. In the conformal theory such functions are
normally zero since the vortex operators have the dimension one.  But
in the string theory we calculate the averages of a type $< \int d^2 z
\hat V_n(z)>$ integrated over the parameterization space. They are
already quantities of zero dimension, and the formal integration leads
to the ambiguity $0*\infty$ which should in general give a finite
result. Another possible reason for vanishing of the one-point
correlator could be the additional infinite W-symmetry found in
\cite{FAT} for the CFT (\ref{LSG}) (at $R=3/2, \ \tmu=0$) \footnote{We
thank A.Zamolodchikov for this comment}. The generators of this
symmetry do not commute with the vortex operators $\hat V_n, \ n\ne\pm
1$. Hence its vacuum average should be zero, unless there is a singlet
component under this symmetry in it. Our results suggest
the presence of such a singlet component.

Note that the one-point functions were calculated \cite{FB} in a
non-conformal field theory with the Lagrangian:
\be
 L={1\over4\pi}\left[ (\partial x)^2 +(\partial\phi)^2
 -4\hat R\phi 
  +m \left( e^{-\hf\phi} +e^{\hf\phi}\right)\cos({3\over 2} x)\right].
\label{NCFT}
\ee
This theory coincides with the CFT (\ref{LSG}) (again at $R=3/2$ and 
$\tmu=0$) in the limit $m\to 0$, $\phi_0\to -\infty$, 
with $\lambda=m e^{-\hf\phi_0}$
fixed, where $\phi_0$ is a shift of the zero mode of the Liouville
field $\phi(z)$.
So we can try to perform this limit in the calculated one-point
functions directly. As a result we obtain 
$\K_n^{(m)} \sim m^2 \lambda^{3n-4}$. 
The coefficient we omitted is given by a complicated
integral which we cannot perform explicitly. It is important
that it does not depend on the couplings and is purely numerical.
We see that, remarkably, 
the vanishing mass parameter enters in a constant power
which is tempting to associate with the measure $d^2 z$ of
integration. Moreover, the scaling in $\lambda$ is the same as
in (\ref{bh1point}) (at $R=3/2$) up to the $n$-independent factor
$\lambda^{-8}$. All these $n$-independent factors disappear if we consider the
correlators normalized with respect to $\K_1^{(m)}$ 
which is definitely nonzero. 
They behave like $\sim \lambda^{3(n-1)}$ what coincides with
the MQM result.

The two point correlators for the CFT (\ref{LSG}) with $\tmu=0$ are
calculated in \cite{FB} only in the case of vortex operators of
opposite and equal by modulo vorticities. They should be in principle
compared to our $\K_{n,-n}$ correlators from (\ref{pm2pc}). However,
we should compare only properly normalized quantities, knowing that
the matrix model correlators generally differ from the CFT correlators
by the leg-factors.  To fix the normalization we have to compare the
quantities similar to (\ref{pm2pcn}) and (\ref{pp2pcn}) which are not
available in the CFT approach.

A lot of interesting physical information is contained in multipoint
correlators (3-point and more) which we did not consider in this paper
and some of which are calculated in the CFT approach of
\cite{FZZ,TESCHNER}. The direct calculations from the Hirota equations
look very tedious.  Hopefully the collective field theory approach of
\cite{KOS} can help on this way.

A lot is to be done to learn how to extract information about the
black hole physics from the correlators. First of all, we have only
calculated some vortex (winding modes) correlators in our MQM
approach.  The vertex (momentum modes, or tachyon) correlators are
unavailable in the Toda hierarchy approach used here and do not seem
to fit any known integrable structure in the MQM, although they are in
principle calculable directly from the MQM in the large $N$ limit. On
the other hand, they bear even a more important information about the
system than the winding correlators: they describe the $S$-matrix of
scattering of tachyons off the black hole and can be used to ``see''
the black hole background explicitly. The one-point tachyon
correlators may give information about the black hole radiation.  In
general, the MQM approach seems to be a good chance to work out a
truly microscopic picture for the 2D black hole physics.

\section*{Acknowledgments}

We would like to thank V. Fateev, I. Kostov, D. Kutasov 
and  A. and Al. Zamolodchikov for interesting discussions. 
This research of S.A. was supported in part by European network 
EUROGRID HPRN-CT-1999-00161.
The research of V.K. was supported in part by European Contract
HPRN-CT-2000-00122.

\appendix

\section*{Appendix A}
Let us investigate which terms will survive in the dispersionless
limit in the equation (\ref{eq1}). 
This limit corresponds to $\xi\longrightarrow 0$.
It is clear that all terms have the
form
\be
\left[\prod\limits_{j=1}^l D_{k_j}\right] \tau_{s+1}\cdot \tau_s
\label{term1} 
\ee
with $n=\sum\limits_{j=1}^l k_j$ fixed. When we rewrite them in terms
of the free energy $F_s=\log\tau_s$ we obtain a set of terms of a kind
\be
e^{F_s+F_{s+1}}\prod\limits_{j=1}^p \left(\left[\prod\limits_{i=1}^{q_j} 
\frac{\partial}{\partial t_{k_{j_i}}}\right] (F_{s+1}+(-1)^{q_j}F_s) \right)
, \label{term2}
\ee
where $\sum\limits_{j=1}^p q_j=l$. 
In the dispersionless limit (\ref{fe}), (\ref{yxi}) and (\ref{tau-f})
imply
\beq 
F_{s+1}+ F_s &\sim & 2F_0, \\
F_{s+1}- F_s &\sim & -i\xi\frac{\partial}{\partial \w}F_0,
\eeq
The main contribution in the dispersionless limit comes from the terms
with the minimal degree of $\xi$. The total degree of $\xi$ for a
given term is 
\be 
\sum\limits_{j=1}^l \Delta[t_{k_j}]-2p+r=l-2p+r-\frac{nR}{2},
\ee
where $r$ is the number of odd $q_j$.
It is easy to see that the minimum is achieved when
all $q_j=1$ or $2$ 
and it equals $-\frac{nR}{2}$ what is independent
on all particular parameters and hence is 
similar for all terms in the equation. 
It means that only two- and one-point correlators 
survive.

\section*{Appendix B}
Combine together all terms with $\tD_k$ and separately with
$\tX_k$ in the exponents of the eq. (\ref{eq3}). Then 
due to vanishing of $\tD^2\tX_k$ we can apply the formula 
$e^{A+B}=e^Ae^Be^{-\frac12[A,B]}$
which is valid for operators having c-number commutator. 
This gives
\beq 
&
\frac{2}{x-y}\left( x e^{\sum\limits_{k=1}^{\infty}x^k \tX_k}
e^{\frac12 \sum\limits_{k,l=1}^{\infty}x^{k+l} \tX_{k,l}}
- y e^{\sum\limits_{k=1}^{\infty}y^k \tX_k}
e^{\frac12 \sum\limits_{k,l=1}^{\infty}y^{k+l} \tX_{k,l}} \right) = &
\nonumber \\
 & \left( 1+\sum\limits_{m=1}^{\infty} y^m (D_{m}+X_m)\right) 
 e^{\sum\limits_{k=1}^{\infty}x^k \tX_k}
e^{\frac12 \sum\limits_{k,l=1}^{\infty}x^{k+l} \tX_{k,l}} +
\left( 1+\sum\limits_{n=1}^{\infty} x^n(D_{n}+X_n) \right)
e^{\sum\limits_{k=1}^{\infty}y^k \tX_k}
e^{\frac12 \sum\limits_{k,l=1}^{\infty}y^{k+l} \tX_{k,l}} & \label{man1}
\eeq
The right hand side of this equation can be rewritten as follows
\beq
 &\left(1+ \sum\limits_{m=1}^{\infty}m y^m \left(\tX_m+
 \sum\limits_{n=1}^{\infty}x^n \tX_{n,m}\right) \right)
 e^{\sum\limits_{n=1}^{\infty}x^n \tX_n +
 \frac12 \sum\limits_{k,l=1}^{\infty}x^{k+l} \tX_{k,l}} +
& \nonumber \\
 &
\left( 1+\sum\limits_{n=1}^{\infty}n x^n\left( \tX_n +
\sum\limits_{m=1}^{\infty}y^m \tX_{n,m}\right) \right) 
e^{\sum\limits_{m=1}^{\infty}y^m \tX_m +
\frac12 \sum\limits_{k,l=1}^{\infty}y^{k+l} \tX_{k,l}}. 
& \label{man2}
\eeq
Due to (\ref{X0n}) we obtain
\beq 
&
\frac{x+y}{x-y}\left( 
e^{\frac12 \sum\limits_{k,l=0}^{\infty}x^{k+l} \tX_{k,l}}
-  e^{\frac12 \sum\limits_{k,l=0}^{\infty}y^{k+l} \tX_{k,l}} \right) =&
\nonumber \\
 & \left(
 \sum\limits_{n,m=0}^{\infty}m x^n y^m \tX_{n,m}\right) 
 e^{\frac12 \sum\limits_{k,l=0}^{\infty}x^{k+l} \tX_{k,l}} +
\left(\sum\limits_{n,m=0}^{\infty}n x^n y^m \tX_{n,m}\right) 
e^{\frac12 \sum\limits_{k,l=0}^{\infty}y^{k+l} \tX_{k,l}} =0, &
\label{man3}
\eeq
what can be written in terms of the generating function (\ref{genf})
as in (\ref{eqMain}).

\end{document}